\documentclass{aastex}
\usepackage{graphicx,emulateapj5}
\def\gsim{\;\rlap{\lower 2.5pt
 \hbox{$\sim$}}\raise 1.5pt\hbox{$>$}\;}
\def\lsim{\;\rlap{\lower 2.5pt
   \hbox{$\sim$}}\raise 1.5pt\hbox{$<$}\;}

\def\myputfigure#1#2#3#4#5%
{\vskip#5pt\makebox[0pt]{\hskip#2in
\includegraphics[width=#3\textwidth]{#1}}\vskip#4pt\hfill}

\submitted{to appear in {\it The Astrophysical Journal Letters}}

\def\aap{A\&A}
\def\apjl{ApJ}
\def\apj{{\em Ap. J. }}

\def\mnras{{\em Mon. Not. Roy. Ast. Soc. } }

\def\om{$\Omega_m$}
\def\ol{$\Omega_\Lambda$}
\def\sig{$\sigma_8$}

\begin{document}

\title{Constraints on \om, \ol, and \sig\, from
Galaxy Cluster Redshift Distributions}
\author{Gilbert Holder$^1$, Zolt\'an Haiman$^{2,}$\altaffilmark{3}, Joseph J. Mohr$^4$ } 
\affil{$^1$ Department of Astronomy and Astrophysics, University of 
Chicago, Chicago IL 60637 \\ $^2$ Princeton University Observatory, Princeton
NJ 08544 \\ $^4$ Departments of Astronomy and Physics, University of Illinois,
Urbana IL 61801 }
\email{holder@oddjob.uchicago.edu,haiman@astro.princeton.edu,
jmohr@uiuc.edu}
\altaffiltext{3}{Hubble Fellow}

\begin{abstract}

We show that the 
number and redshifts distribution
of galaxy clusters in future deep cluster surveys can
place strong constraints on the matter density, $\Omega_m$, the vacuum energy
density, $\Omega_\Lambda$, and the normalization of the matter power spectrum,
$\sigma_8$. Degeneracies between these parameters are different from those in
studies of either high--redshift type Ia Supernovae (SNe), or cosmic microwave
background (CMB) anisotropies. Using a mass threshold for
cluster detection expected to be typical for upcoming 
Sunyaev-Zel'dovich effect (SZE) surveys, we find that
constraints on \om\ and \sig\ at the level of roughly 5\% or better can be
expected, assuming redshift information is known at least to $z\sim 0.5$
and in the absence of significant systematic errors.
Without information past this redshift, \ol\ is constrained to $25\%$.  With
complete redshift information, deep ($M_{\rm lim}\sim 10^{14}h^{-1}{\rm
M_\odot}$), relatively small solid angle ($\sim 12\ {\rm deg}^2$) surveys, 
can further
constrain \ol\ to an accuracy of $\sim 15\%$, while large solid angle 
surveys with ground-based, large-format bolometer arrays could measure 
\ol\ to a precision
of $\sim 4\%$ or better.

\end{abstract}


\section{Introduction}

The abundance of galaxy clusters can provide strong constraints on cosmological
parameters (e.g., Bahcall and Fan 1998; Viana and Liddle
1999). \nocite{bahcall98,viana99} Future cluster surveys, especially 
those using the
Sunyaev-Zeldovich effect (SZE), will provide large catalogs of clusters with a
selection function that is remarkably 
simple 
\citep{barbosa96,holder00,kneissl01}.  
Cluster surveys probe the amplitude of the power spectrum as a 
function of redshift and the cosmic volume per unit redshift
and solid angle (Haiman, Mohr and Holder 2000\nocite{haiman00}, hereafter HMH).
This allows unique constraints on cosmology, 
and provides a sensitive test of structure formation.
As we show below,  the cluster redshift distribution (i.e., counts)
out to $z \sim 0.5$ is a powerful probe of 
$\Omega_m$
and $\sigma_8$, while a deeper cluster inventory will allow a 
determination of the vacuum energy
density, $\Omega_\Lambda$, as well.

In HMH we showed expected constraints on the equation of state of the dark
energy and the matter density, assuming a flat universe and two specific
proposed cluster survey instruments.  
In this work, we specialize to the case
of vacuum energy (equation of state $w\equiv dp/d\rho=-1$), but generalize to
non--flat universes; we treat the normalization of the matter
power spectrum $\sigma_8$ as a free parameter.  
Here we use less realistic mass limits for cluster detection, to focus on
the effects of abundance evolution and volume as opposed to the
particulars of the instruments.  
Interpreting a future cluster survey will require careful consideration 
of the survey selection function.

In \S~2, we outline our methods for calculating survey yields, while \S~3
outlines our statistical methods. In \S~4, we present our results and conclude
with a discussion of possible systematic errors and future work that will be
necessary to achieve the expected constraints.

\section{Calculating Cluster Survey Yields}

To calculate the expected number of clusters per square degree as a function of
redshift one must consider several elements: the comoving volume per unit
redshift and unit solid angle, $dV/dz d\Omega$, the minimum observable mass as
a function of redshift and cosmology, $M_{\rm lim}(z)$, and the comoving number
density of halos above the mass threshold as a function of redshift and
cosmology, $n(>M_{\rm lim},z)$.

The comoving volume per unit redshift is straightforward to calculate (e.g.,
Peebles 1993\nocite{peebles93}), and for this work we assume that the mass
threshold is constant with redshift and cosmology, approximately
consistent with expectations for surveys using the Sunyaev--Zeldovich effect
\citep{holder00,bartlett00,kneissl01}. The mass threshold of detectability in a
survey is very important, and it is likely that such a mass threshold will have
a dependence on cosmology.  We ignore the cosmological dependence in this work
in order to isolate the effect of the growth rate of structure as a function of
cosmology as the primary discriminant between cosmological models.  

For the comoving number density of clusters, we use the mass function of dark
matter halos obtained from large cosmological simulations (Jenkins et
al. 2001)\nocite{jenkins01}, where the comoving number density of clusters
between mass $M$ and $M+dM$ is given by
\begin{equation}
{dn \over dM} =  0.315 {\rho_\circ \over M^2} {d \ln \sigma^{-1} \over d\ln M} 
	\exp [-|\ln \sigma^{-1} + 0.61|^{3.8}] 	\quad ,
\end{equation} 
where $\sigma(M,z)$ is the variance in the density field smoothed on a scale
corresponding to mass $M$ and $\rho_\circ$ is the present--day mean density of
the universe.

From linear theory, $\sigma(M,z)$ can be separated into $\sigma(M,z)=
\sigma(M,z=0) D(z)$, where $D(z)$ is the linear growth factor. We calculate
$D(z)$ directly from linear theory (e.g., Peebles 1993\nocite{peebles93}).  The
variance $\sigma(M,z=0)$ is calculated with a top hat filter of radius
appropriate for mass $M$, using the matter power spectrum calculated from the
fitting functions of Eisenstein and Hu (1999). \nocite{eisenstein99a}

As a fiducial model, we take a flat, low-density model, with
$\Omega_m=0.3,\Omega_\Lambda=0.7,h=0.65,\sigma_8=1.0,n=1,$ and $\Omega_b
h^2=0.02$. We keep $h,n$ and $\Omega_b h^2$ fixed for all models and vary the
remaining three parameters.  To isolate the effects of the growth of the power
spectrum, we artificially constrain the power spectrum to be that of the
fiducial model for all models.  In the absence of modifications to the power
spectrum, these results are absolutely unchanged by varying $h$ or $\Omega_b
h^2$.  Relaxing the assumption of a fixed power spectrum will slightly affect
the constraints on parameters, but the qualitative results are unchanged, as
the shape of the power spectrum has only a mild effect on the cluster abundance (HMH).
Large scale structure measurements, from the spatial distribution of
either galaxies or galaxy clusters should be able to provide accurate
measurements of the matter power spectrum over the range of interest.

\section{Estimating Constraints on Cosmology}
\label{sec:stats}

To estimate constraints on cosmological parameters, we assume Poisson errors on
finely binned ($\Delta z=0.01$) model predictions. Still finer binning was 
found to have no effect on the results.  We determine confidence
regions in two ways, using a Monte Carlo method and the Fisher matrix
\citep{eisenstein99}.

In the first method, we construct a likelihood space by a Monte Carlo method,
generating 3000 realizations of the input fiducial model, in the form of mock
catalogs of clusters at different redshifts, and then fitting each
realization,
allowing $\Omega_m, \Omega_\Lambda$ and $\sigma_8$ to find their best-fit
values. In this way, we can map out regions of parameter space that contain
68\% and 95\% of the realizations.

For the case of Gaussian errors, the likelihood is simply related to the usual
$\chi^2$ statistic as $-2 \ln{\mathcal{L}}= \chi^2$. In the case of
Poisson errors, it can be shown that the analogous statistic is the Cash C
statistic \citep{cash79}:
\begin{equation}
-2 \ln \mathcal{L} = -2 \sum_{a=1}^{N} n_a \ln e_a - e_a - \ln n_a ! \quad ,
\end{equation}
where $n_a$ and $e_a$ are the observed and expected number of counts in bin
$a$, respectively, and N is the total number of bins.

A computationally more efficient method to estimate confidence regions is to
use the Fisher matrix.  We constructed the second derivatives of the log
likelihood with respect to the parameters of interest at the position in
likelihood space of the fiducial model.  Ensemble averaging then leads to a
simple expression for the Fisher matrix:
\begin{equation}
{- \partial ^2 \ln \mathcal{L} \over \partial p_i \partial p_j } =
 \sum_{a=1}^{N}
 {\partial e_a \over \partial p_i}
 {\partial e_a \over \partial p_j} {1 \over m_a} \quad ,
\end{equation}
where $m_a$ is the number expected in the fiducial model in bin $a$ and
the $p_i$ are the parameters of interest, in this case \om, \ol, and
\sig.  This is also what one would obtain assuming a Gaussian
probability distribution with $\sigma^2=N$ and using the usual $\chi^2$
estimator for the likelihood.

Assuming that the likelihood distribution is approximately Gaussian near the
peak likelihood, we can use confidence limits for Gaussian statistics (i.e.,
$\chi^2$) to obtain 68\% and 95\% confidence regions.

The advantage of the Fisher matrix method is that it is very fast and 
gives accurate results if the likelihood space is well-behaved.
While we have no reason to suspect this is not the case, 
we perform the Monte Carlo analysis as a useful check. The Fisher matrix
approximates confidence regions as Gaussian ellipsoids, while the true
confidence regions could have broad tails or significant curvature.

\myputfigure{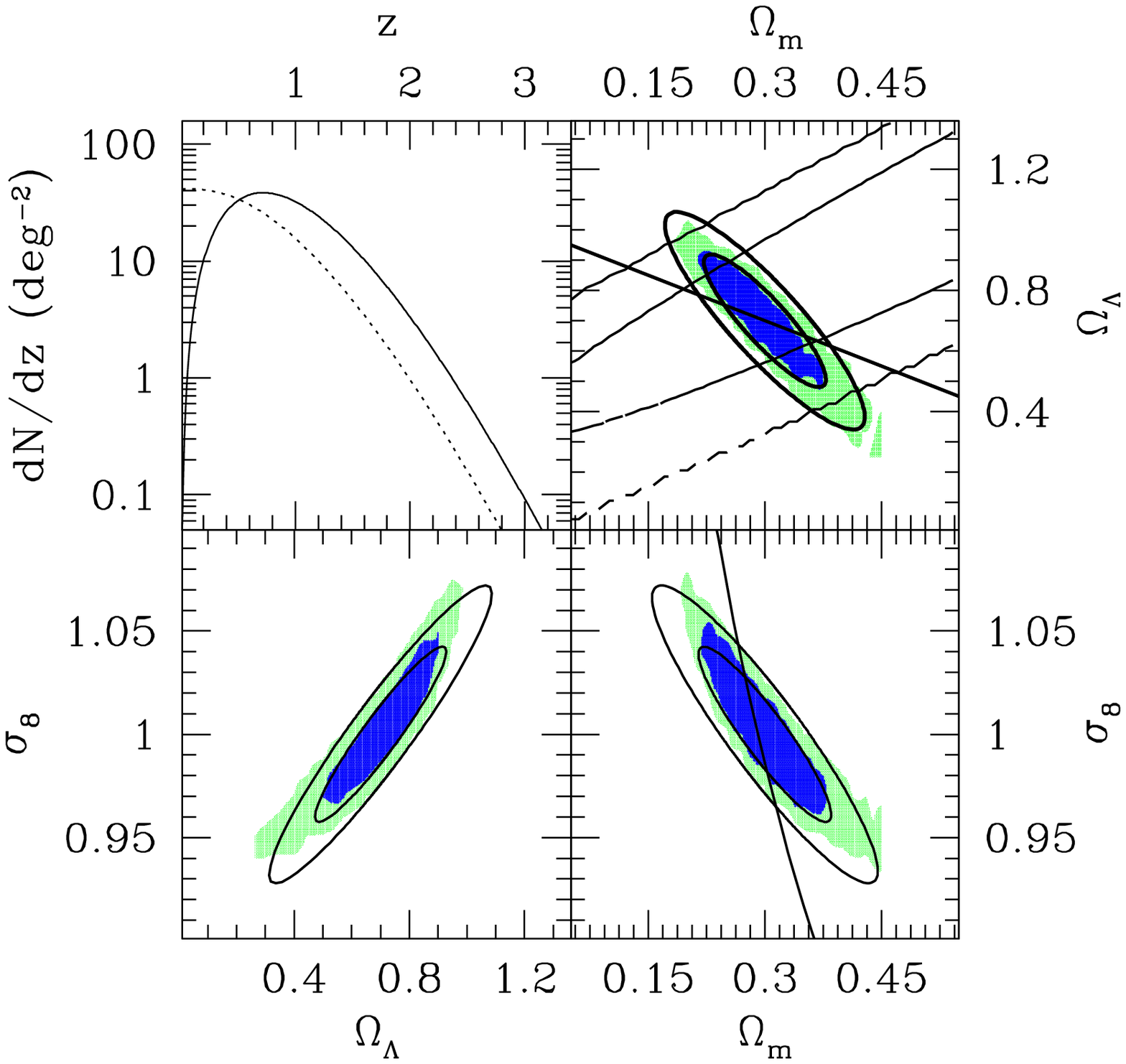}{3.4}{0.57}{-30}{-40}
\figcaption{
Top left panel: differential (solid curve) and cumulative (dashed curve) number
counts of clusters as a function of redshift. A mass limit of $10^{14}
h^{-1}{\rm M_\odot}$ is assumed.  Top right panel:
68\% and 95\% confidence levels on \om-\ol\, marginalized over
\sig. Shaded regions are from a Monte Carlo analysis,
while contours show results using the Fisher matrix.  Dashed contours show the
current constraints from the Supernova Cosmology Project (see
http://supernova.lbl.gov), while the diagonal solid line indicates a flat
universe favored by CMB anisotropies.  Bottom right panel: Confidence levels on
\om-\sig\, marginalized over \om,
similar to the top right panel.  The diagonal line shows the degeneracy from
the abundance of massive 
($M\gsim 2\times 10^{15} {\rm M_\odot}$) local clusters. Bottom left panel:
Confidence levels on \ol-\sig\ (bottom right), marginalized over \om.}

\section{Results}
\label{sec:results}

We first look at a relatively low value for the constant mass limit,
$M_{\rm lim}= 10^{14} h^{-1} {\rm M_\odot}$, appropriate for a deep SZE survey
that would cover approximately 12 ${\rm deg}^2$.
In Figure 1, we show the expected number of clusters as a function
of redshift, as well as the likelihood contours we find in three
different projections in the \om, \ol, \sig\ plane.  
The constraints
from the Monte Carlo method are similar to those obtained from
the Fisher matrix method, especially in the \om-\ol\ plane. 
This indicates that the likelihood space is
well-behaved, with degeneracies manifesting themselves as
ellipses. As the Fisher matrix method is much
faster, this allows a much more extensive study of parameters.
However, the differences are not negligible for the plots involving
\sig, suggesting that the likelihood space in the \sig\ direction is
significantly non-Gaussian.

Constraints on \om\ and \ol\ from cluster surveys are complementary to other
probes. Figure 1 also shows the current constraints on parameters from studies
of distant supernovae, as well as the \om+\ol=1 line expected from inflation
and preferred by CMB studies
\citep{debernardis00,hanany00,netterfield01,pryke01}.
The different orientations of the 
parameter degeneracies suggest that joint constraints from CMB measurements,
supernovae, and deep cluster surveys will be a powerful probe of both
cosmological parameters and potential systematic effects in other data sets.
In addition, Figure 1 shows that the degeneracy in the \om-\sig\ plane is in a
different direction when compared with local determinations from massive
($M\gsim 2\times 10^{15} {\rm h^{-1}~M_\odot}$) clusters \citep{viana99}. 
This is due to the different mass and redshift ranges probed, with the 
current local determinations coming from much more massive clusters (HMH).

When marginalized over the other two parameters, we can expect uncertainties in
\om, \ol, and \sig\ of 6\%, 15\%, and 3\%, respectively, assuming complete
redshift information.  The relatively small number of ($\sim 400$) clusters
spread over only a few square degrees should allow follow up observations to
get redshift information.

\myputfigure{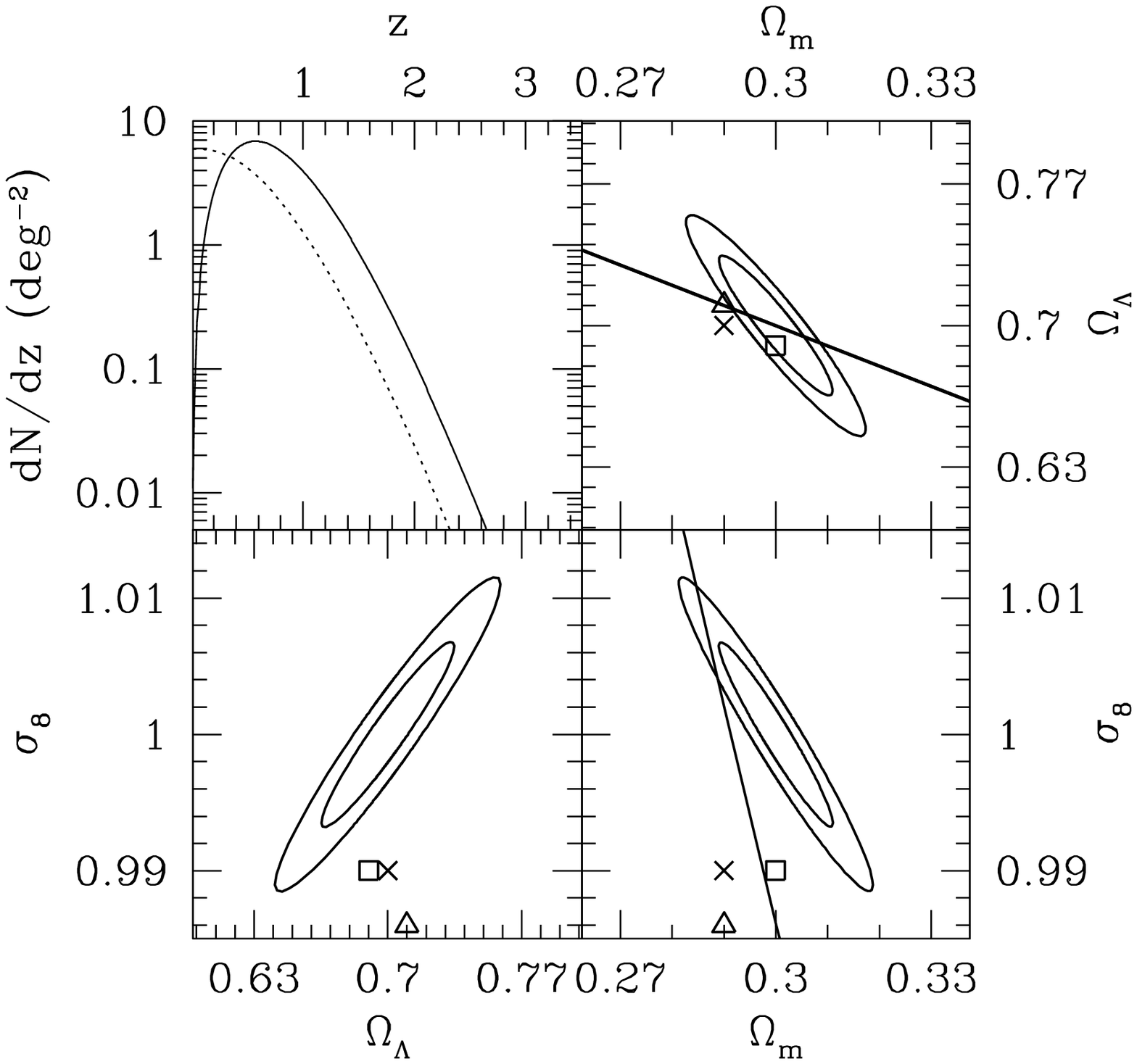}{3.6}{0.55}{-30}{-40}
\figcaption{Confidence levels on parameters, 
marginalized over the third parameter
in the likelihood space, as in Figure 1, but for a higher mass limit of
$2.5\times10^{14}h^{-1}{\rm M_\odot}$.  Contours show the expected $68\%$ and
$95\%$ confidence regions, using the Fisher matrix. Relative axis scaling 
and the labeling of the curves is the same as Figure 1, except in the 
upper right panel, where we have omitted showing the constraints
from SNe. Squares, triangles, and X marks show results of Monte Carlo
investigations of the effects of a systematic tilt in 
the mass function (x $\sigma^{0.1})$, 
a systematic reduction in the mass function amplitude ($-10\%$) and 
a systematic offset in the limiting mass ($+5\%$), respectively.
\vskip5pt}

Large ground-based telescopes equipped with bolometer arrays will be efficient
at detecting clusters through their SZE signatures. We assume a typical mass limit
for such instruments to be $2.5\times 10^{14} h^{-1} {\rm M_\odot}$ and a
typical sky coverage of 4000 ${\rm deg}^2$.  We again calculate constraints on
cosmological parameters using the Fisher matrix, with results shown in Figure
2. In this case, the possible constraints on cosmology are approximately
an order of magnitude stronger than the first case. 
The
direction of the degeneracy is slightly different from Figure 1, due to the
different redshift distribution and mass range. This indicates the potential
power of using the distribution of observed clusters in mass, which we have not
utilized in this paper, but will explore in future work. The 
offset between the local \om-\sig\  determination and our central value is an 
artifact of our
assumed fiducial model being slightly offset from the best fit of
Viana and Liddle (1999), although our fiducial model is well within their quoted
uncertainties.

Single parameter uncertainties on \om, \ol, and \sig\ are expected to be
approximately 0.7\%, 2\%, and 0.5\% respectively, for this case, assuming
complete redshift information.  Again, redshift information will be very
important for such tests.  In this case, redshifts for clusters spread over
4000 ${\rm deg}^2$ is a non-trivial task, but it is clear from Figure 2 that
the reward for such an exercise will be precise measurements of cosmological
parameters.

Also shown in Figure 2 are the effects on parameter estimation of possible
systematic effects. The Monte Carlo method was used to find the best-fit cosmology
when the fiducial model was subject to systematic errors.
The effects of
uncertainty in the true mass function were treated as either a small tilt,
where the mass function has been multiplied by $\sigma(M)^{0.1}$, or as
an overall normalization error, where we reduced the amplitude of the mass function
by $10\%$.  Such effects are allowed within the uncertainties of \cite{jenkins01}. 
The final systematic error that we tested was an uncertainty of $5\%$ in the
determination of the limiting mass. This is significantly smaller than current
uncertainties in mass estimates, but could be achievable with high signal-to-noise
X--ray or SZE observations of galaxy clusters.
Systematic errors could affect
determinations of $\sigma_8$, with biases in \om\ and \ol\ roughly 
comparable to the $1\sigma$ statistical uncertainty.  

We finally adopt a relatively high value for the limiting mass, $M_{\rm lim}=
8\times10^{14} h^{-1} {\rm M_\odot}$, appropriate for an SZE survey like
{\it Planck}, covering a large fraction of the sky (we assume 50\%).  
From Figure 3,
it can be seen that such a survey would allow strong constraints on 
cosmological parameters, comparable to those expected from the second
case above.
Even with only
$\sim 1$ cluster per square degree, the large solid angle allows a large
catalog to be assembled.

In this case, redshift determination would be challenging.  
A reasonable strategy would be to cross-correlate with a large 
solid angle
survey such as the SDSS.  For example, it is reasonable to expect that
every cluster found by {\it Planck} with $z<0.5$ will appear in the 
SDSS catalog in regions of overlapping coverage.  
Using only those clusters with $z<0.5$, and
again assuming 50\% of the sky is covered (although SDSS will only cover $\sim
25\%$), Figure 3 shows the expected constraints on parameters. The constraints
on \sig\ and \om\ are still strong, but with only the nearby sample it is
difficult to constrain the vacuum energy.  
At lower redshifts, increasing \ol\ results in an increased volume 
that largely offsets the effect of the growth function.
At least in the case that we are considering where the mass limits are largely
decoupled from cosmology, 
the clusters past $z\sim 0.5$ are most useful
in constraining the cosmological constant.
This is unfortunate, as these are the clusters that will be 
most difficult to follow up. 

Single parameter uncertainties on \om, \ol\ and \sig, assuming complete
redshift information are 1\%, 4\%, and 0.9\%, respectively for a survey like
{\it Planck}.  For redshift information only out to $z=0.5$, these numbers become
2\%, 26\%, and 2\%, respectively. 
The changes in \om\ and \sig\ uncertainties
are barely larger than would be expected from the increased
Poisson error 
from the smaller number of clusters. For a ground-based
bolometer array, the same effect can be observed, with expected uncertainties
of 3\%, 23\%, and 3\%, respectively, without redshift information past $z=0.5$.

\myputfigure{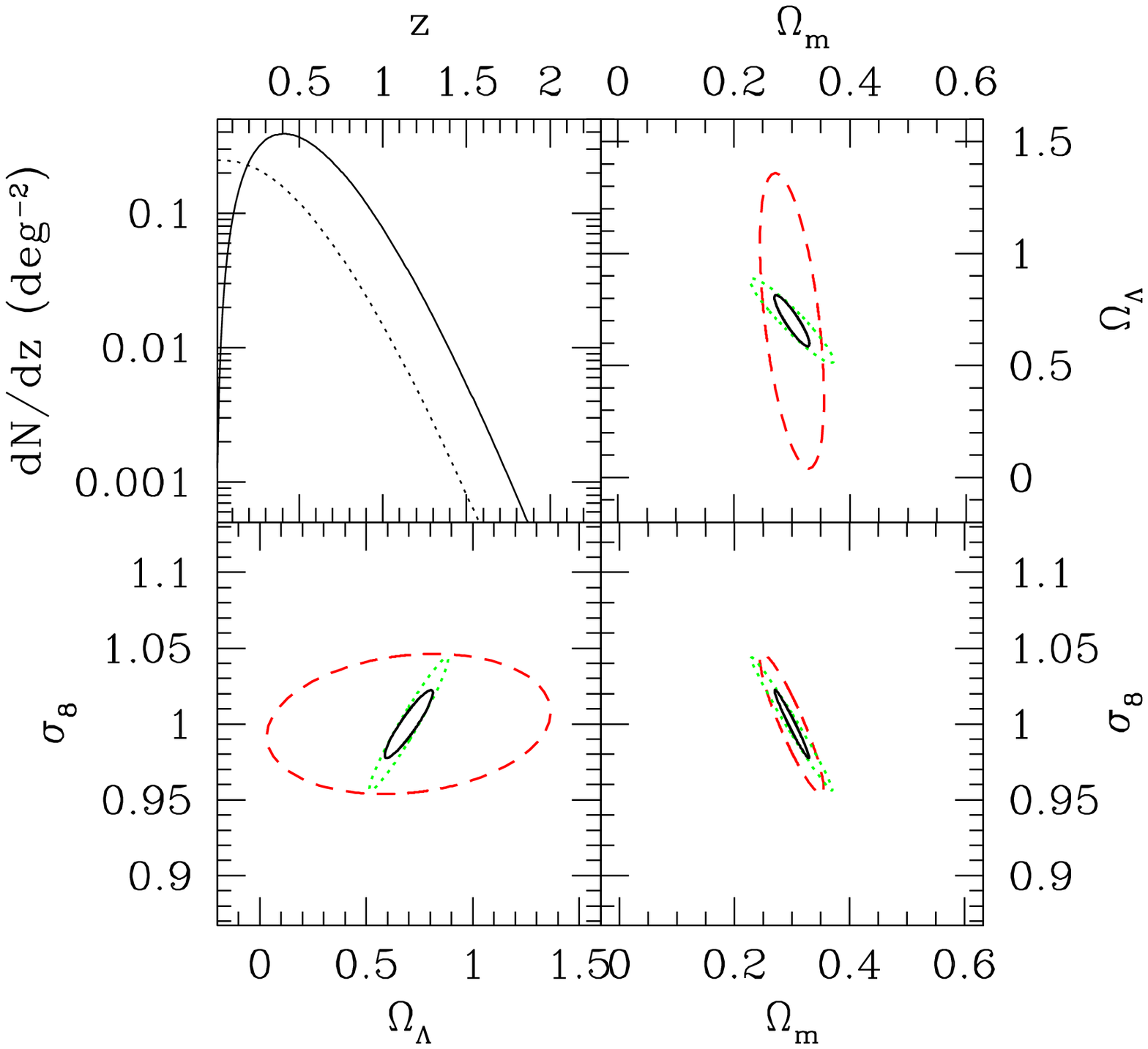}{3.6}{0.55}{-30}{-40}
\figcaption{Confidence levels on parameters, using the Fisher matrix as
in Figure 2, but for a mass limit of $8\times10^{14}h^{-1}{\rm M_\odot}$.
Dashed ellipses show the expected $95\%$ confidence regions using only clusters
with $z<0.5$, dotted curves show the $95\%$ confidence regions using only
clusters with $z>0.5$, while solid curves show the $95\%$ confidence region for
the entire sample. Relative axis scaling is the same as Figure 1.}

\section{Conclusions and Future Work}

We have shown that cluster surveys have the potential to be powerful probes of
cosmological parameters, and in a manner that will serve as a useful check of
other methods. Efficient follow-up to obtain redshift information will be
crucial, with most information on \ol\ coming from clusters with $z>0.5$.  This
argument would favor a deep survey on a relatively small patch of sky, as
follow-up observations could be done on a smaller scale.

The results obtained here are based on highly idealized toy models, and are far
from realistic in several respects.  
Most importantly, we have not addressed here the effects of systematic errors
from cluster evolution, which can mimic different values of cosmological
parameters (but see HMH for a quantitative discussion).  In addition, although
we expect the matter power spectrum to cause relatively little cosmological
sensitivity (from HMH), this deserves further study 
We assume that Poisson errors are setting the uncertainty on cosmological
parameters. With surveys in the design stage, it is difficult to estimate the
potential systematic errors that will be important in any interpretation of
real surveys.

The
mass function, in principle, carries a significant amount of information on
cosmology and could also be a source of systematic error in an interpretation
of survey results.  The mass function has been accurately (to an overall error
of 30 percent) determined in a handful of cosmologies by Jenkins et al. (2001).
It will be important to repeat similar numerical experiments in a broader range
of cosmologies, to have a better understanding of the scaling of the mass
function with cosmology.
Taking full advantage of the statistical power in the surveys 
considered here will
require a description of the mass function, which is better than
$10\%$ accurate.

Mass limits as a function of cosmology must be well-understood to interpret the
source counts. This includes both evolution with redshift and scatter in the
``detectability'' at a given mass.  Although temperatures can be obtained for
clusters in a cosmologically independent way from their X--ray spectra,
the relation of this temperature to the halo mass in different cosmologies
needs further study.  Extracting unbiased cosmological parameters from real
survey data will require multi-wavelength follow-up of at least a sub-sample of
the detected clusters in order to understand the mass limit of the survey 
to better than $5\%$.

Just as large scale structure surveys have learned about galaxy evolution and
CMB experiments have learned about foregrounds, cluster surveys will learn
about galaxy cluster structure and evolution. As the largest virialized objects
in the universe, clusters at high redshift will yield a wealth of information
on galaxy formation and other aspects of cosmology.  A study of cluster scaling
relations that combines SZE and X-ray data can help in separating the
systematic effects arising from the assumption of virialization from cosmology
\cite{verde01}.

None of these issues should, in principle, prevent robust and highly precise
constraints on cosmological parameters from cluster surveys.  The justified use
of the Fisher matrix will enable fast exploration of many cosmological
parameters. The small subset explored here and in HMH suggest that these
constraints should be highly complementary to large scale structure studies
\nocite{phillips00}
(e.g., Phillips et al. 2000), CMB anisotropies, and probes of distance, such as
type Ia supernovae.  In addition, tracing the redshift evolution of the cluster
abundance offers a unique opportunity to track the growth of structure that is
expected from hierarchical clustering, providing a strong test of the
underlying cold dark matter model.

\acknowledgements{
We acknowledge many useful
discussions with John Carlstrom, Michael Turner and David Spergel.
GPH was supported in part by a DOE grant to the University of Chicago.
ZH is supported by NASA
through the Hubble Fellowship grant HF-01119.01-99A, awarded by the Space
Telescope Science Institute, which is operated by the Association of
Universities for Research in Astronomy, Inc., for NASA under contract NAS
5-26555. JJM is supported in part by grants G01-2125X and AR1-2002X 
awarded by the Chandra Science Center.  The Chandra Science Center
is operated by the Smithsonian Astrophysical Observatory for NASA
under contract NAS8-39073.}

\bibliographystyle{apj}

\end{document}